\documentstyle[emulateapj,apjfonts,epsf]{article}

\def\ltsima{$\; \buildrel < \over \sim \;$} 
\def\gtsima{$\; \buildrel > \over \sim \;$} 
\def\lsim{\lower.5ex\hbox{\ltsima}} 
\def\gsim{\lower.5ex\hbox{\gtsima}} 
\def\msun{${\rm M_\odot}$} 
 
\begin{document} 
 
\title{Where May Ultra-Fast Rotating Neutron Stars Be Hidden?}
 
\author{Luciano Burderi,\altaffilmark{1} 
Andrea Possenti,\altaffilmark{2} 
Francesca D'Antona,\altaffilmark{1} 
Tiziana Di Salvo,\altaffilmark{3}
Marta Burgay,\altaffilmark{4} 
Luigi Stella,\altaffilmark{1} 
Maria Teresa Menna,\altaffilmark{1} 
Rosario Iaria,\altaffilmark{5}
Sergio Campana,\altaffilmark{6} 
Nichi d'Amico\altaffilmark{2}} 
\medskip 
 
\altaffiltext{1}{Osservatorio Astronomico di Monteporzio,  
via Frascati 33, 00127 Roma, Italy} 
\altaffiltext{2}{Osservatorio Astronomico di Bologna,  
via Ranzani 1, 40127 Bologna, Italy} 
\altaffiltext{3}{Astronomical Institute "Anton Pannekoek," University of 
Amsterdam and Center for High-Energy Astrophysics,
Kruislaan 403, NL 1098 SJ Amsterdam, the Netherlands.} 
\altaffiltext{4}{Universit\`a di Bologna, Dipartimento di Astronomia,  
via Ranzani 1, 40127 Bologna, Italy} 
\altaffiltext{5}{Dipartimento di Scienze Fisiche ed Astronomiche,
Universit\`a di Palermo, via Archirafi 36, 90123 Palermo, Italy} 
\altaffiltext{6}{Osservatorio Astronomico di Brera,  
via Bianchi 46, 23807 Merate, Italy} 
 
\bigskip 
 
\begin{abstract} 
 
The existence of ultra-fast rotating neutron stars (spin period $P \la 1$
ms) is expected on the basis of current models for the secular evolution of
interacting binaries, though they
have not been detected yet. Their formation depends on the quantity of matter
accreted by the neutron star which, in turn, is limited by the mechanism of
mass ejection from the binary. An efficient mass ejection can avoid the
formation of ultra-fast pulsars or their accretion induced collapse to a
black hole.
We propose that significant reductions of the mass-transfer rate may cause 
the switch-on of a radio pulsar phase, whose radiation pressure
may be capable of ejecting out of the system most of the matter transferred
by the companion. This can prevent, {\it for long orbital periods and if a
sufficiently fast spin has been reached}, any further accretion,
even if the original transfer rate is restored, thus limiting the minimum 
spin period attainable by the neutron star.
We show that close systems (orbital periods $P_{\rm orb} \sim
1$ hr) are the only possible hosts for ultra-fast spinning neutron
stars. This could explain why ultra-fast radio pulsars have not been
detected so far, as the detection of pulsars with very short spin 
periods in close systems is hampered, in current radio surveys, by strong 
Doppler modulation and computational limitations.
 
\end{abstract} 


\section{Introduction} 
 
The shortest spin period ever observed for a rotating neutron star
(hereafter NS), $P_{\rm min}=1.56$ ms (Backer et al. 1982), is not a 
physical limit for the stability of spun-up NSs. In fact, the period
$P_{\rm min}$ is longer than the limiting period, $P_{\rm lim}$, below
which the star becomes unstable to mass shedding at its equator.
Cook, Shapiro \& Teukolsky (1994) have shown that  
$P_{\rm lim} \sim 0.5$ ms for most equations of state (EoS) adopted to 
describe the ultra-dense nuclear matter.

The re-acceleration
of a NS to ultra-short periods depends remarkably on the amount 
of mass (and hence of angular momentum) accreted. 
Typically $M_{1 ms}\sim 0.35$ \msun\ must be accreted to reach $P\sim 1$ ms
(e.g.\ Burderi et al. 1999). 
Most donor stars in systems hosting recycled millisecond radio pulsars (MSP)
have certainly lost, during their interacting binary evolution, a mass $\ga
M_{1 ms}$. They now appear as white dwarfs of mass $\sim 0.15-0.30$ \msun\
(e.g.\ Taam, King \& Ritter 2000), whose progenitors are likely to have been
stars of $\sim 1.0-2.0$ \msun\ (Webbink, Rappaport \& Savonije 1983; Burderi,
King \& Wynn 1996, Tauris \& Savonije 1999). Examples of possible
evolutionary sequences, in which the assumption of {\it conservative} mass
transfer is plausible, are computed in \S 2. Hence, in order to explain 
the lack of ultra-fast rotations, we have to find physical mechanisms able
to prevent the accretion of a considerable fraction of the mass lost by the 
companion ($M_{lost}=0.8-1.7$ \msun) onto the NS.
 
In this Letter we propose a mass ejection mechanism based on the  
sweeping effects of the energy outflow from a rapidly spinning NS
undergoing a radio pulsar phase.  
The sweeping effects of the energy outflow from a  
rapidly spinning NS have been firmly established since  
the early works of Shvartsman (1970) and Illarionov \& Sunyaev (1975).
Ruderman, Shaham \& Tavani (1989) and Shaham \& Tavani (1991)  
discussed in particular the case of recycled pulsars in low mass binary  
systems. As soon as the accreting plasma moves out beyond the light-cylinder  
radius (where an object corotating with the NS attains the speed of light, 
$R_{\rm LC} = c P / 2 \pi$), the NS becomes generator of magnetodipole  
radiation and relativistic particles, whose pressure may expel the matter
overflowing the Roche lobe (see also Campana et al. 1998).   
In the following we determine the dependence of this mass ejection mechanism  
on the parameters of the system (\S 3) and show that it  
naturally provides both an explanation for the values of the mass and  
rotation of the observed MSPs (Thorsett \& Chakrabarty 1999; 
Tauris \& Savonije 1999) and an indication about  
where ultra-fast spinning NSs could reside (\S 4), suggesting 
in turn why they could have been elusive up to now. 
 
\section{Accretion and Ejection during Mass Transfer}
 
It is well known that an upper limit to the accretion rate is given by
the Eddington limit.
However, for typical initial masses $\lsim 1.6$ \msun\
and initial orbital periods $P_{orb} \lsim 10$ days, the donor transfers  
mass at {\it sub}-Eddington rates, in principle making all $M_{lost}$    
available for the recycling of the NS. 
We explored the conservative mass transfer scenario computing 
the system evolution (with initial and final parameters listed in
Table~\ref{tab:dantona})
with the ATON1.2 code (D'Antona, Mazzitelli, \& Ritter 1989).
The mass loss rate is computed following the formulation by
Ritter (1988), as an exponential function of the distance of the stellar
radius to the Roche lobe, in units of the pressure scale height.
This method also allows to compute the first phases of mass transfer, during
which the rate reaches values which can be much larger than the stationary
values, due to the thermal response of the star to mass loss. For systems
having a mass donor of 1.2\msun\ we assumed either that the donor fills the
Roche lobe while it is evolving towards the red giant branch (case B) or
during the core-hydrogen burning phase (case A).
Results are shown in Table~1.

In cases 2, 3 and 4, mass transfer is guided by the thermal and nuclear
evolution of the secondary; in case 1 it is guided by magnetic braking. 
The latter evolution is expected to stop at $P_{\rm orb} \simeq 2.5$ hr, when
the star becomes fully convective. It may resume at a shorter period
guided by gravitational wave loss of angular momentum, like in the
cataclysmic binary systems, reaching a minimum period of $\sim 1.03$ hr.
For the sequences 2, 3 and 4, the final period is in the range 90 -- 150 hr;
the final mass of the donor is in the range $0.21-0.33$ \msun\ (see
Tab.~\ref{tab:dantona}), in agreement with the masses of the companion stars
of MSPs as inferred from accurate timing measures (e.g. Burderi, King, \&
Wynn 1998).
On the other hand the assumption of conservative mass transfer
leads to final NS masses in the range $2.3-2.7$ \msun, much
larger than the inferred masses of the NSs in these systems and dangerously
close to (or even larger than) the maximum mass allowed for a NS in most of 
the proposed EoS (Cook, Shapiro \& Teukolsky 1994). 
This suggests that either the mass transfer cannot be conservative 
or that the final NS must be very massive implying that 
accretion-induced collapse to a black hole is a likely outcome for LMXBs. 
Indeed the masses of a sample of radio pulsars in binary systems 
were estimated by Thorsett \& Chakrabarty (1999). These are 
consistent with a remarkably narrow Gaussian 
distribution, with $M_{\rm NS}=1.35 \pm 0.04~{\rm M}_\odot$, although
the sample is contaminated by five relativistic NS-NS binary
systems, the progenitors of which are massive X-ray binaries. 
If the proposed value of the amount of mass accreted, $<0.1~{\rm  M}_\odot$, 
is representative of the MSP observed sample, this suggests that significant 
mass losses occur during the Roche lobe overflow phase.

A propeller effect (Illarionov \& Sunyaev 1975) has been often invoked
to explain this discrepancy. In the widely accepted scenario of 
accretion onto a magnetized NS (see {\it e.g.} Hayakawa 1985 
for a review) the accretion disc is truncated at the magnetospheric 
radius $R_{\rm m}$, at which the magnetic field pressure equals the 
pressure of the matter in the disc.  
The value of $R_{\rm m}$\ for a Shakura--Sunyaev accretion disc
({\it e.g.} Burderi et al. 1998) is, in
a first approximation, close to $\sim 0.5$ times the Alfven radius $R_A$:
$R_m \simeq 0.5 \times R_A=1.1 \times 10^6 \mu_{26}^{4/7} R_6^{-2/7}
L_{37}^{-2/7} m^{1/7} {\rm cm}$,
where
$L_{37}$ is the accretion luminosity in units of $10^{37}$ erg/s, 
$m$ is the NS mass in solar masses, 
$R_6$ is the NS radius in units of $10^6$ cm, 
$\mu_{26}$ is the magnetic moment of the NS in units of 
$10^{26}$ G cm$^3$ ($\mu = B_{\rm s} R^3$ with $R$ and $B_{\rm s}$
NS radius and surface magnetic field along the
magnetic axis, respectively). 
Accretion onto a spinning magnetized 
NS is centrifugally inhibited once $R_{\rm m}$ lies outside 
the corotation radius $R_{\rm CO}$, the radius at which the Keplerian 
angular frequency of the orbiting matter is equal to the NS 
spin: 
$ 
R_{\rm CO} = 1.50 \times 10^{6} m^{1/3} P_{-3}^{2/3} \;\; {\rm cm},  
\label{eq:rco} 
$ 
where $P_{-3}$ is the spin period in milliseconds.    
In this case a significant fraction of the accreting matter could in
principle be ejected from the system: this is called propeller phase.

The virial theorem sets stringent limits on the fraction of matter that can
be ejected in this phase. In fact, it states that, at any radius in the
disc, the virialized matter has already liberated ({\it via} electromagnetic
radiation) half of its available energy.
Considering that $R_{\rm m} \sim R_{\rm NS} $ for
MSPs, the matter at the magnetosphere has radiated $\sim 50\%$ of
the whole specific energy, $E_{\rm acc} = GM / R_{\rm NS}$, 
obtainable from accretion.  To eject the same matter (close to the
NS surface) $\sim 1/2 ~E_{\rm acc} $ must be given back to it.  As
the only source of energy is that stored in the NS by the
accretion process itself, the typical
ejection efficiency is $\sim 50 \%$.  Actually, with a fine tuned
alternation of accretion and propeller phases of the right duration,
the ejection efficiency can be higher than $50\%$,
although a difficulty with this scenario is that,
once the system has reached the spin--equilibrium, no further
spin up takes place and the storage of accretion energy in a
form that allows its subsequent re--usage for ejection is
impossible.  
Thus the accreted mass is about a half of the transferred mass, i.e.  
$\simeq 0.4-0.8$ \msun. This has two main consequences:
a) even taking a propeller phase into account, NSs in MSPs will either be 
very massive or even collapse into black holes, 
b) as the amount of mass accreted is considerably
larger than the minimum required to spin up the NSs to ultra short
periods, one has to invoke an {\it ad hoc} final,
long lasting propeller phase with a highly effective spin down
to form the observed population of moderately fast spinning MSPs. 

An alternative viable hypothesis to explain the lack of ultra-fast rotating
NS is that gravitational wave emission balances the torque due to accretion
(see a review in Ushomirsky, Bildsten, \& Cutler 2000).
However, while these emission mechanisms, whose importance in LMXBs will be 
probably clarified in the next future by detectors such as LIGO-II, might
explain the observed spin periods of spun-up NSs, which seem to cluster 
between 260 and 590 Hz (see e.g.\ van der Klis 2000), they cannot solve
the problem of the large final masses of NSs. 

The only way to overcome these difficulties is to obtain ejection
efficiencies close to unity. This is indeed possible if the
matter is ejected so far away from the NS surface that it has an almost
negligible binding energy $GM / r \ll E_{\rm acc}$. As the NS is spinning
very fast, the switch on of a radio pulsar is unavoidable once $R_{\rm m} \ge
R_{\rm LC}$.
The pressure exerted by the radiation field of the radio pulsar may overcome
the pressure of the accretion disc, thus determining ejection of 
matter from the system. Once the disc has been swept away,
the radiation pressure stops the infalling matter as it overflows the inner
Lagrangian point, where $E_{\rm acc} \sim 0$.
 
\section{The Effect of the Pulsar Energy Outflow} 
 
The push on the accretion flow exerted by the (assumed dipolar) magnetic 
field of the NS can be described in terms of an outward pressure  
(we use the expressions {\it outward} or {\it inward pressures} 
to indicate the direction of the force with respect to the radial direction):  
$P_{\rm MAG} = B^2/4\pi = 7.96 \times 10^{14} \mu_{26}^2 r_6^{-6} \;\; 
{\rm dy}/{\rm cm}^2 
\label{eq:pmag} $,
where $r_6$ is the distance from the NS center in units of $10^6$ cm. 
If the disk terminates outside $R_{\rm LC}$, the outward 
pressure is the radiation pressure of the rotating 
magnetic dipole, which, assuming isotropic emission, is 
$P_{\rm PSR} = 2.04 \times 10^{12} P_{-3}^{-4}  
\mu_{26}^2 r_6^{-2} \;\; {\rm dy}/{\rm cm}^2. 
\label{eq:ppsr} $
In Figure~1 the two outward pressures ($P_{\rm MAG}$ and $P_{\rm PSR}$) 
are shown as bold lines for typical values of the parameters 
(see figure caption). 
 
The flow, in turn, exerts an inward pressure on the field. 
For a Shakura--Sunyaev accretion disc
(see {\it e.g.} Frank, King \& Raine 1992): 
$P_{\rm DISK} = 1.02 \times 10^{16} \alpha^{-9/10}  
n_{0.615}^{-1} L_{37}^{17/20} m^{1/40} R_6^{17/20} f^{17/5} r_6^{-21/8}  
\;\; {\rm dy}/{\rm cm}^2  
\label{eq:pgas}  $, 
where $\alpha$ is the Shakura--Sunyaev viscosity parameter,
$n_{0.615} = n/0.615 \sim 1$ for a gas with solar abundances
(where $n$ is the mean particle mass in units of the proton mass),
and $f = [1 - (R_6/r_6)^{1/2}]^{1/4} \le 1$. 
As before we measure $\dot M$ in units of $L_{37}$, supposing that 
all the matter reaches the NS surface.  
Note that this equation is valid in the zone C of the dik, that is 
at radii larger than $3.45 \times 10^{7}$ cm for a luminosity of 
$10^{37}$ erg s$^{-1}$ (Burderi, King, \& Szuszkiewicz 1998). 
In Figure~1 the inward disc pressure  
for a luminosity $L_{\rm MAX}$, corresponding to a high accretion rate, 
is shown as a thin solid line. 
The disc pressure line, which intersects $P_{\rm MAG}$ at $R_{\rm LC}$, 
defines a critical mass-transfer
rate $M_{\rm switch}$ ({\it i.e.} a critical luminosity $L_{\rm
switch}$, shown as a dashed line in Fig.~1) at which the 
radio pulsar switches-on. 

The intersections of the $P_{\rm DISC}$ line corresponding to $L_{\rm MAX}$ 
with each of the
outward pressure lines define equilibrium points between the inward and outward
pressures. The equilibrium is {\it stable} at $r = R_{\rm m}$, and {\it
unstable} at $r = R_{\rm STOP}$, which can be derived equating $P_{\rm PSR}$
and $P_{\rm DISK}$:
\begin{eqnarray} 
R_{\rm STOP} 
           \sim 8 \times 10^{11} \alpha^{-36/25}
n_{0.615}^{-8/5} R_6^{34/25} f^{136/25} &  &  \nonumber \\
L_{37}^{34/25} m^{1/25} \mu_{26}^{-16/5} P_{-3}^{32/5}  \;\; {\rm cm}.  & &
\label{eq:rstop} 
\end{eqnarray} 
In fact, as $P_{\rm MAG}$ is steeper than $P_{\rm DISC}$, if a
small fluctuation forces the inner rim of the disc inward (outward),
in a region where the magnetic pressure is greater (smaller) than the
disc pressure, this results in a net force that pushes the disc back
to its original location $R_{\rm m}$. As $P_{\rm PSR}$ is flatter than 
$P_{\rm DISC}$, with the same argument is easy to see that no stable 
equilibrium is possible at $R_{\rm STOP}$ and the disc is swept away 
by the radiation pressure. 
This means that, for $r>R_{\rm STOP}$, no disc can exist for
any luminosity $\leq L_{\rm MAX}$.

It is convenient to divide the systems into ``compact" and ``wide"
depending whether the primary Roche-lobe radius ($R_{\rm L 1}$)
lies inward or outward $R_{\rm STOP}$, as
they behave very differently in response to significant 
variations of the mass-transfer rate. 
The dependence of $R_{\rm L 1}$ on the orbital parameters is given by:
$$
R_{\rm L 1} = 3.5 \times 10^{10} P_{\rm h}^{2/3}
(m + m_2)^{1/3} \left[ 1 - 0.462 \left( {m_2} \over {m+m_2} 
\right)^{1/3} \right] \;\; {\rm cm}
\label{eq:rl1} 
$$
using the approximation given by Paczynski (1971), 
where $P_{\rm h}$ is the orbital period in hours and $m$, $m_2$ are
the NS and the companion masses in \msun, respectively.
Therefore, the $P_{\rm orb}$ that separates ``compact" and ``wide" systems
can be obtained imposing $R_{\rm STOP}/ R_{\rm L 1} = 1$ and solving for
$P_{\rm orb}$:
\begin{eqnarray}
P_{\rm crit} =  1.05 \times (\alpha^{-36}  
n_{0.615}^{-40} R_6^{34})^{3/50} L_{36}^{51/25} m^{1/10} \mu_{26}^{-24/5} 
P_{-3}^{48/5}  &  &  \nonumber \\
\left[ 1 - 0.462 \left( {m_2} \over {m+m_2} 
\right)^{1/3} \right]^{-3/2} 
(m+m_2)^{-1/2}
\;\; {\rm h}  &  & 
\label{eq:pcrit} 
\end{eqnarray}
where $L_{36}$ is $L_{\rm MAX}$ in units of $10^{36}$ erg/s.
(In the following, always keep in mind that the separation between
``compact" and ``wide" systems depends on $P_{-3}^{48/5}$, and has no
`absolute' meaning).

When the luminosity alternates between its maximum value $L_{\rm MAX}$ 
and a minimum luminosity $L_{\rm MIN} < L_{\rm switch}$ 
the behavior of a ``compact" system is cyclic.
During the high state, the magnetospheric radius 
is smaller than both the corotation and the
light--cylinder radius and the NS will normally accrete matter and
angular momentum, thus increasing its spin (accretion phase).
The sudden drop in the mass-transfer rate to $L_{\rm MIN}$
initiates a phase, that we termed ``radio
ejection'', in which the mechanism that drives {\it mass overflow}
through $L_1$ is still active, while the pulsar
radiation pressure prevents mass accretion. 
As the matter released from the companion cannot accrete, it is now
ejected as soon as it enters the Roche lobe of the primary. 
When the system goes back to the high state, the disk might form again
and the accretion can resume. 

The response of a ``wide" system to the same variations of the accretion rate
is quite different. When the mass-transfer rate recovers the value
corresponding to the high state, the cyclic behavior is lost. Indeed, since
in this case $R_{\rm L 1}$ is located beyond $R_{\rm STOP}$, $P_{\rm DISC} <
P_{\rm PSR}$ at $R_{\rm L 1}$ even in the high state and the accretion cannot
resume. This means that for a ``wide" system, once a drop of the mass-transfer
rate has started the radio ejection, a subsequent restoration of the original
mass-transfer rate is unable to quench the ejection process (as already
pointed out by Ruderman, Shaham, \& Tavani 1989).
 
In conclusion, while the evolution without a radio ejection phase
implies that a large fraction of the transferred mass is accreted onto
the NS (because of the constraints imposed by the virial theorem), 
we have demonstrated that the switch-on of a radio
pulsar (associated to a significant drop in mass transfer) could
determine ejection efficiencies close to $100\%$, as the matter is
ejected before it falls into the deep gravitational potential well of
the primary. 
For ``compact" systems, we have shown that
radio ejection is swiftly quenched by a resumption of the
original mass-transfer rate, leading to the prediction that the amount
of mass accreted is substantial. On the other hand, if a radio
ejection starts in a ``wide" system, this implies that the
accretion is inhibited in the subsequent evolution.

\section{Where to Search Ultra-Fast Spinning NS} 
 
A statistical analysis based on the current samples of detected  
MSPs (Cordes \& Chernoff 1997) proved that, using different  
hypotheses for the period distributions of these sources, there is always  
a non negligible probability for periods $P<P_{min}$.  
Possenti et al. (1999, 1998) performed population synthesis calculations  
including propeller and randomly choosing the mass accreted onto the NS  
in the interval $0.01 - 0.4$ \msun\ (this corresponds to efficiency  
of ejection between $60$\% and $99$\%). They confirmed that the process of  
recycling in low-mass binaries can produce a significant amount of  
ultra-rapidly rotating objects, under different assumptions for the evolution  
of the mass transfer and the magnetic field. However, despite the large  
efforts devoted in the last years (D'Amico 2000; Crawford, Kaspi \& Bell 2000;
Edwards et al. 2001) no pulsar with $P<P_{min}$ has been observed so far. 
 
The more rapid the NS spins, the smaller is the drop of ${\dot M}$ needed 
to switch the pulsar on and expel the accretion disc (see \S 3). 
Once this occurred a new phase of accretion 
is possible only if the orbital period is short enough.  
Eq.~(\ref{eq:pcrit}) predicts that spinning a NS up to $P\lsim 1$~ms requires  
very close ($P_{orb}\lsim 1.5$ hrs) binary systems.  
 
The orbital Doppler shift on the pulsar signal can provide  
a natural observational bias against the detection of ultra-rapidly spinning  
pulsars. All the codes for searching pulsations from a source in a close  
binary system are a compromise between computational capability and  
sensitivity: on each data set, they must perform a two dimensional search in  
the space of the unknown parameters dispersion measure (DM, related 
to the distance of the object) and acceleration (resulting from binary 
motion). Short data sets reduce the Doppler period shifts during the 
observation and relax the computational requests, but at the price of  
limiting the sensitivity (see e.g. Camilo et al. 2000). As a consequence,  
up to now orbital periods shorter than $\sim 90$ min have been poorly  
searched, even in the more favorable case of targeted searches  
(as those pointing to globular clusters), where one of the two parameters  
(DM) is known. 
This observational bias could be enhanced in presence of eclipses (favoured 
in very close binary pulsar systems, Nice et al. 2000) or in case of 
a large duty cycle of the pulsar signal, a low radio luminosity and a strong  
interstellar scintillation, already suggested for the elusiveness of  
ultra fast rotating pulsars (Possenti 2000).  
 
Even in the favorable case $P_{\rm orb} <P_{\rm crit}$, 
a steady accretion during all the Roche lobe overflow 
phase (i.e.\ no ejection episodes) could prevent the  
formation of a ultra-fast spinning NSs: in fact if the EoS is soft
the NS could undergo a collapse to black hole.  
Actually, Cook, Shapiro \& Teukolsky (1994) showed that (because of the 
centrifugal pull determined by the rapid NS rotation)  
a NS can accrete matter even beyond the limit at which the black-hole 
collapse takes place. 
However, when the accretion halts and the fast spinning NS 
loses rotational energy via magnetodipole emission, a radial instability 
can set in leading to the formation of a black-hole (see Stella \& Vietri 
1999).  
 
In summary we have shown that: 
1. For very fast rotating NSs, if variabilities in the matter flow trigger an
extended radio-pulsar phase, accretion onto a NS cannot recover unless also
the orbital period is quite short. Thus extremely compact binary systems are
strongly favoured for harbouring ultra fast rotating pulsars. 
2. If the equation of state for the nuclear matter is soft enough,
binary evolution with steady accretion rate would often lead to  
black-hole formation, via accretion induced collapse. 
 
\acknowledgments 
This work was partially supported by a grant from the Italian Ministry
of University and Research (Cofin-99-02-02).


\newpage 
 
\begin{deluxetable}{lcccc} 
\tablecaption{\label{tab:dantona} Mass Evolution in Low Mass Binaries. }
\tablecolumns{5} 
\tablehead{\colhead{$~$} 
&\colhead{System type}&\colhead{${M_{NS}}({\rm M_\odot})$}
&\colhead{${M_{\it donor}}({\rm M_\odot})$}&\colhead{${P_{orb}}({\rm hrs})$}} 
\startdata 
Initial & 1       & 1.40 & 1.20 & 10.46 \\
Final   & 	  & 2.58 & 0.02 & 1.42  \\
        &         &      &      &       \\   
Initial & 2       & 1.40 & 1.20 & 30.55 \\
Final   & 	  & 2.35 & 0.25 & 338.1 \\ 
        &         &      &      &       \\   
Initial & 3       & 1.40 & 1.40 & 33.95 \\
Final   & 	  & 2.54 & 0.26 & 483.8 \\ 
        &         &      &      &       \\   
Initial & 4       & 1.40 & 1.60 & 34.00 \\
Final   & 	  & 2.67 & 0.33 & 346.0 \\ 
\tableline 
\enddata 
\end{deluxetable} 

 
\begin{figure}
\plotone{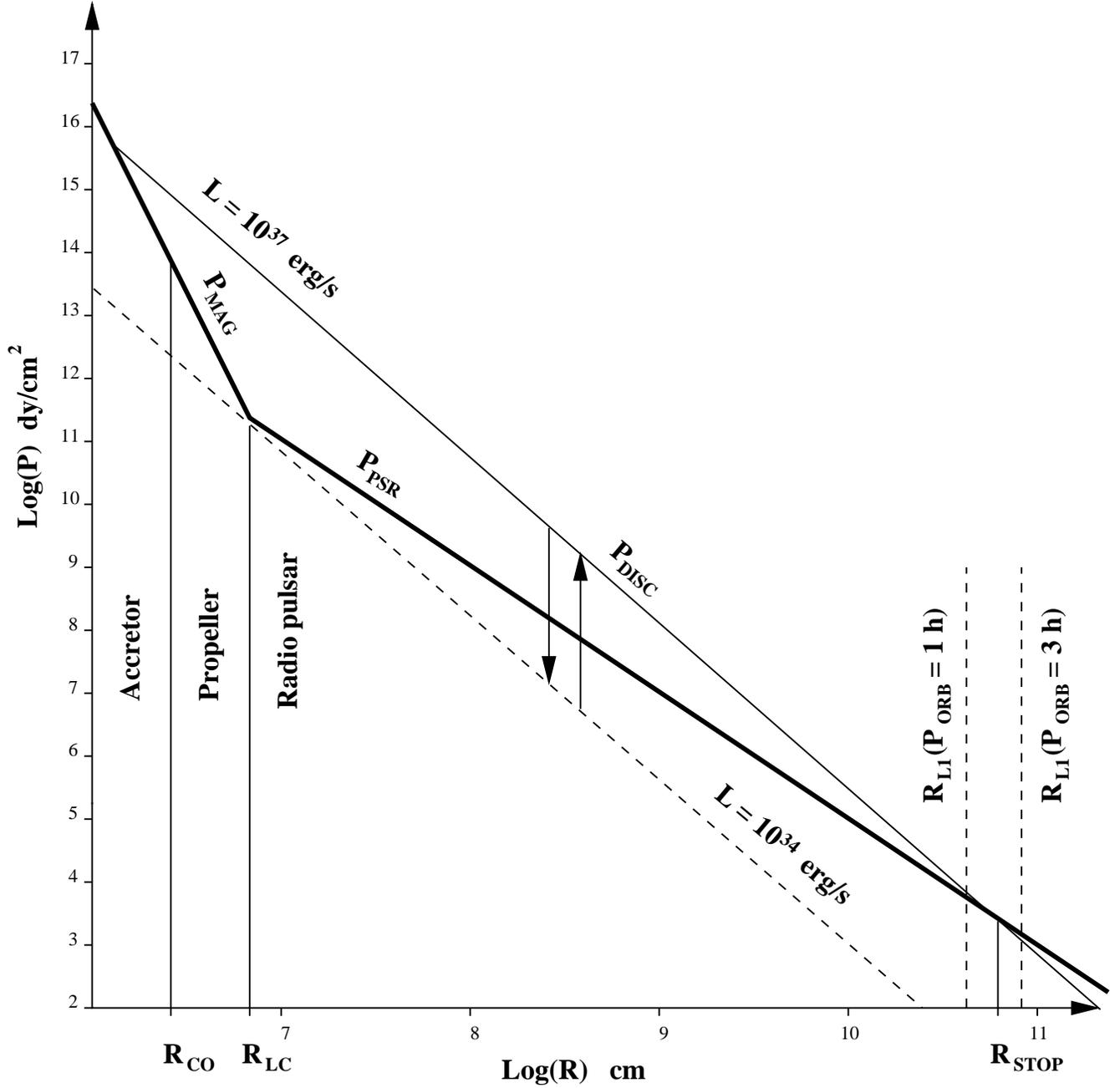}
\caption{\label{fig:fig1} The radial dependence of the pressures relevant for the 
evolution of accreting NSs and recycled pulsars. The parameters adopted are: 
$\mu_{26} = 5$, $P_{-3} = 1.5$, $\alpha = 1$, 
$n_{0.615} =1$, $R_6 = 1$, $m = 1.4$, $f = 1$.}
\end{figure}

\end{document}